# Deleting object selective units in a fully-connected layer of deep convolutional networks improves classification performance

Yuta Kanda[1], Kota S Sasaki[1, 2], Izumi Ohzawa[1, 2], Hiroshi Tamura[1, 2]

[1]Graduate School of Frontier Biosciences, Osaka University, Suita, Osaka 565-0871, Japan

[2]Center for Information and Neural Networks, Suita, Osaka 565-0871, Japan

**Corresponding author:** Dr. Hiroshi Tamura

Laboratory for Cognitive Neuroscience, Graduate School of Frontier Biosciences, Osaka University, 2A1, Center for Information and Neural Networks, 1-4 Yamadaoka, Suita, Osaka 565-0871, Japan

Tel: +81-6-6879-7969; Fax: +81-6-6879-4439; E-mail: tamura@fbs.osaka-u.ac.jp

**Number of pages**: 18; **Number of Figures**: 6; **Number of Tables**: 1

**Acknowledgements**: We thank Edanz Group (https://en-author-services.edanzgroup.com/) for editing a draft of this manuscript. This work was supported by a Grant-in-Aid for Scientific Research on Innovative Areas: Innovative SHITSUKAN Science and Technology (No. JP15H05921) from MEXT, Japan. The authors declare no competing financial interests.

Neurons in the primate visual cortices show a wide range of stimulus selectivity. Some neurons respond to only a small fraction of stimulus images, whereas others respond to many stimulus images in a non-selective manner. It is unclear how stimulus selective and non-selective neurons contribute to visual object recognition. Herein, we examined the relationship between stimulus selectivity and the effect of deletion of units on task performance using fully a connected layer of two types of deep convolutional neural networks (DCNNs). Deleting a stimulus selective unit caused slight improvements of task performance, whereas deleting stimulus non-selective units caused a significant decrease in task performance. However, these findings do not imply that stimulus selective units have no use for the task. Indeed, better performance was obtained when the networks consisted of both stimulus selective and non-selective units.

**Introduction**

Stimulus selective units have been found in the primate brain including humans. In the inferior temporal cortex of monkeys, some neurons respond selectively to an image of objects (Tanaka, 1996; Gross, 2000). However, not all neurons in the visual cortices show selective responses. Rather, some neurons show broad or stimulus non-selective responses (Tamura and Tanaka, 1991). Although these non-selective neurons may not contribute to stimulus encoding (Olshausen and Field 2006), some studies suggest the importance of non-selective neurons for visual object recognition (Leavitt et al. 2017; Zylberberg 2018).

Although stimulus selective neurons are likely to play important roles in visual-object recognition tasks, their significance in the task has never been tested directly. Herein, we examined the relationship between stimulus selectivity and the effect of deletion of a unit on task performance using deep convolutional neural networks (DCNNs). Because classification performance of DCNNs is similar or even better than that of humans (Russakovsky et al., 2015), and units in the higher layers of DCNNs respond selectively to stimulus images (Zeiler and Fergus 2014), DCNNs are a good model for hierarchically-organized primate visual cortices.

We found a negative relationship between stimulus selectivity and changes in task performance; i.e., deleting stimulus non-selective units resulted in a significant decrease in the task performance, while deleting a stimulus selective unit resulted in a slight improvement of task performance. However, these findings do not imply that stimulus selective units have no use for the task. Indeed, better performance was obtained when the networks consisted of both stimulus selective and non-selective units.

**Methods**

Analysis was performed with Alexnet (Krizhevsky et al., 2012) and VGG-19 (Simonyan et al., 2014) software installed on a computer running Windows 10 pro (Microsoft, Redmond, WA, USA). Outputs from the DCNNs were examined with MATLAB (Mathworks, Natick, MA, USA). Alexnet contains five convolutional layers, three pooling layers, two normalization layers, three fully connected layers, and an output-softmax layer. VGG-19 contains 16 convolutional layers, five pooling layers, three fully connected layers, and an output-softmax layer. A rectified linear operation was applied to the output of convolutional layers and two of the fully connected layers (fc6 and fc7). Both Alexnet and VGG-19 were pretrained for classification of 1,000 object categories using the ImageNet database (Deng et al., 2009)

For the analysis of task performance of the DCNNs, we selected images from a validation set of the Imagenet Large Scale Visual Recognition Challenge (ILSVRC) 2012, which included 50,000 images (50 for each of the 1,000 categories). Because the set included images that cannot be recognized well by DCNNs, we selected five images with the best performance for each of the 1,000 categories. Performance of the DCNNs to the selected 5,000 images and that to the original 50,000 images are shown in Table

1.

We deleted a unit by decreasing the weight (the connection strength between units in the preceding layer and the deleting unit) and the bias of the unit (the constant activation level of the deleting unit) to zero, because outputs of a unit were calculated by multiplying the outputs of the preceding layer with weights and adding the bias.

In the present study, we examined effects of deletion using two fully connected layers (fc6 and fc7), which correspond to the higher visual cortical areas of primates. Each of the layers contains 4,096 units. To examine the contribution of each unit for the classification task, we deleted one unit and compared the performance of one-unit deleted DCNN with that of the original DCNN. We calculated the 'loss', which represents the cross entropy and the distance between the DCNN output vector and the teacher vector (Krizhevsky et al., 2012; Simonyan et al., 2014).

$$loss = -\sum_{i=1}^{5000} L_i \log S_i$$

Here, $i$ is the index of images, and $L_i$ and $S_i$ are the vectors of the ground truth of the image and the output vector of DCNNs to $i$-th image, respectively. The smaller the loss, the better the performance. For example, if the deletion of a unit resulted in an increase in the loss, the performance of the DCNN decreased by the deletion and the unit contributed to the task performance of the DCNN. By contrast, if the deletion of a unit did not change the loss, the performance of the DCNN is not affected by the deletion and the unit did not contribute to the task performance of the DCNN. The effect of unit deletion on the loss was evaluated with the loss_ratio.

$$loss\_ratio = \frac{loss_{deletion} - loss_{original}}{loss_{original}} \times 100$$

The stimulus selectivity of the unit was evaluated with the category selectivity index (CSI), which is similar to a common measure used in primate neurophysiology (De Valois et al. 1982).

$$CSI = \frac{R_{max} - R_{others}}{R_{max} + R_{others}}$$

$R_{max}$ and $R_{others}$ represent the response to the stimulus that evoked the largest response and the average across responses to all the other stimuli, respectively. Here, the response is the average across five responses to five stimuli in a category. Therefore, this index is termed the CSI. If the CSI is closer to one, the unit responded to only one category, while if CSI is closer to zero, the unit responded to many categories. We also used sparseness as a measure for stimulus selectivity (Vinje and Gallant, 2000).

$$sparseness = \left(1 - \frac{(\sum r_k)^2}{N \sum r_k^2}\right) \Bigg/ \left(1 - \frac{1}{N}\right)$$

Here, N is the number of stimuli (i.e., 5,000), while $r_k$ is the response to k-th stimulus image. If the sparseness is closer to one, the unit responded to only one stimulus, while if the sparseness is closer to zero, the unit responded to many stimuli.

For statistical analyses, all the data were pooled. Statistical testing was performed using MATLAB (The MathWorks, MA). Statistical threshold ($P$ value) was set at 0.01. If $P$ value was < 0.001, we described this result as $P$ < 0.001, instead of providing exact $P$ value. We also provided test statistics (correlation coefficient or z-value).

**Results**

*Stimulus selectivity of units in the higher layers of DCNNs*

Units in the DCNNs showed a range of stimulus selectivity. Some units in the DCNNs were activated by only a few stimuli (CSI = 0.96, sparseness = 0.95, fc6 of Alexnet; Fig. 1A top), while others were activated by many stimuli (CSI = 0.71, sparseness = 0.76, fc6 of Alexnet; Fig. 1A bottom). The median CSI of fc6 and fc7 using Alexnet was 0.90 and 0.91, respectively (Fig 1B). The median sparseness of fc6 and fc7 using Alexnet was 0.90 0.90, respectively (Fig. 1C). We compared CSI and sparseness across layers of the DCNNs, and found that units in convolutional layer 1 (conv1)–conv4 of Alexnet had much lower CSI and sparseness than those in the higher layers. In other words, the relatively sharp stimulus selectivity of fc6 and fc7 is a characteristic of higher layer units in DCNNs (Fig. 1D).

*Effect of deletion of a single unit in higher layers of DCNNs on the classification performance*

Deletion of a single unit in the fully connected layers of DCNNs induced changes in loss. The deletion of the selective unit in Figure 1A decreased the loss (loss_ratio =

−0.051), meaning that deletion improved the performance. The deletion of the non-selective unit in Figure 1A increased the loss (loss_ratio = 0.293), meaning that deletion impared the performance. The median loss_ratio was 0.053 for fc6 and 0.064 for fc7 of Alexnet (Fig. 2A). These values were significantly different from zero (fc6: $P < 0.001$, z-value = 41.4; fc7: $P < 0.001$, z-value = 54.9; Wilcoxon signed rank test). The median loss_ratio of fc6 and fc7 of VGG-19 was also positive and different from zero (fc6: loss_ratio = 0.023, $P < 0.001$, z-value = 33.4; fc7: loss_ratio = 0.030, $P < 0.001$, z-value = 53.5; Fig. 2B). These data suggest that deletion of a single unit impaired performance of the categorization task on average. Note that deletion of some units improved the task performance, because deletion of some units (e.g., the unit in Figure 1A top) decreased the loss.

The degree of stimulus selectivity of the fully-connected layers was negatively correlated with the loss_ratio. In the fc6 and fc7 of Alexnet, the correlation coefficients between CSI and the loss_ratio were −0.43 (Spearman's correlation coefficient, $P < 0.001$, test of independence from zero) and −0.09 ($P < 0.001$), respectively (Fig. 3A). In the fc6 and fc7 of VGG-19, the correlation coefficients between CSI and the loss_ratio were −0.32 ($P < 0.001$) and −0.10 ($P < 0.001$), respectively (Fig. 3B). Similar results were observed for the relationship between sparseness and the loss_ratio. In the fc6 and fc7 of Alexnet, correlation coefficients between sparseness and the loss_ratio were −0.51 ($P < 0.001$) and −0.08 ($P < 0.001$), respectively. In the fc6 and fc7 of VGG-19, correlation coefficients between sparseness and the loss_ratio were −0.41 ($P < 0.001$) and −0.13 ($P < 0.001$), respectively. Especially in the fc6, both CSI and sparseness were significantly and negatively correlated with the loss_ratio, indicating that deletion of the stimulus non-selective unit causes a significant decrease in classification-task performance in fc6. More importantly, deletion of some of the stimulus-selective units in fc6 caused a negative loss_ratio, indicating that their deletion improved classification-task performance.

The negative correlation between stimulus selectivity and the loss_ratio was also observed in the other layers of the DCNN. Thus, we examined the relationship in the convolutional layers. In all the convolutional layers except for conv5 of Alexnet and conv1_2 and conv5_4 of VGG-19, we found a similar negative correlation (Fig. 4), suggesting that this is a general tendency across layers of the DCNNs.

Overall, these data suggest that the selective units are not important or are even deleterious for the performance of a classification task in all categories. Alternatively, selective units may only contribute to the classification of a related stimulus category, but are deleterious for the other categories. Thus, deletion may result in overall improvements. To examine these hypotheses, we calculated the loss_ratio for the related category with the stimulus selective and deleterious units in fc6 of Alexnet, which has a CSI larger than the median and a negative loss_ratio (n = 597). The loss_ratio for the related category of the selected units (i.e., stimulus selective and deleterious) was $1.96 \times 10^{-4}$ (median) and was larger than zero ($P < 0.001$, z-value = 16.6, Wilcoxon signed rank test), while that for the other categories was $-1.38 \times 10^{-4}$ and was less than zero ($P < 0.001$, z-value = 21.2, Wilcoxon signed rank test; Fig. 5A). The same trend was observed in the fc6 of VGG-19 (n = 621; Fig. 5B). Thus, deletion of stimulus selective units decreased the performance of the corresponding category, but improved the performance of the other categories, suggesting that stimulus selective units only contribute to the classification of the related stimulus category, but are deleterious for the other categories.

*Effect of deletion of multiple units in higher layers of DCNNs on the classification performance*

The deletion of a stimulus-selective single unit improved the performance of the DCNNs. These data suggest that units with relatively higher stimulus selectivity may not be necessary, and that networks consisting of units with relatively less selective units may have better performance. To directly test this hypothesis, we examined the relationships between the performance and the ratio of stimulus-selective units and non-selective units.

In a fully connected layer, we deleted multiple units simultaneously, and examined the relationship between classification performance and the ratio of non-selective units. Here, selective units were defined as units with a CSI larger than the median, while the other units were defined as non-selective. As a measure of classification performance, we used the Top-5 accuracy rather than loss. In the previous analyses where only a single unit was deleted, the loss_ratio was approximately 0.5%, while in the present analyses where multiple units were deleted, the loss ratio was >1000%. Large-values of loss_ratio may not be sensitive enough for the quantification of changes induced by changes in the ratio of selective units. Thus, for the present analysis, we used Top-5 accuracy, which is based on the relative order, and only changes mildly with deletion of multiple units.

The best performance was obtained if the networks consisted of both stimulus selective and non-selective units (Fig. 6). In other words, only networks with selective or non-selective units showed worse performance. The same results were obtained with sparseness as a measure of the stimulus selectivity, and with different thresholds for stimulus selectivity.

**Discussion**

In the present study, we examined the contributions of stimulus selective units, which respond to only a limited range of stimuli, as well as non-selective units, which respond to many more stimuli, to the classification performance of DCNNs. We found that the deletion of stimulus non-selective units resulted in a larger decrease in the classification performance of DCNNs, suggesting that the non-selective units contributed to the performance. Stimulus selective units contributed to the classification of the responsive category, but not to the other categories. As a result, the overall contributions of selective units were small. More importantly, deletion of some of the selective units improved the overall performance. These results were obtained with two types of DCNNs (Alexnet and VGG-19), confirming that the findings are not specific to a particular type of DCNN, but rather are general to DCNNs trained for object categorization. Further, we obtained similar results with different measures of stimulus selectivity. Overall, these findings suggest that stimulus selective units are unnecessary for task performance. However, we found that the best performance was obtained with DCNNs that included both stimulus selective and non-selective units.

The majority of previous neurophysiological studies have examined stimulus selective units (Tanaka, 1996; Gross, 2000). These selective units are likely to encode information about responding stimulus. In the present study, we found that selective units in DCNNs indeed contributed to the classification performance of responding stimulus. However, because of their sharp stimulus selectivity, these units did not contribute, or were even deleterious, to the encoding of the other stimulus images. Stimulus non-selective units are generally considered unimportant for object recognition or categorization, and are thought to encode more basic components of the stimulus images. However, in the present study, we clarified the importance of non-selective units for object-categorization task. Nevertheless, it remains unclear why the non-selective units play a more important role in the task, and whether the same responses are observed in the primate brain. We would like to point that those non-selective units showed relatively broad stimulus selectivity, but they did not respond to all the images at the same activation level. Therefore, it might be better to call these units as relatively non-selective.

A previous study examining the effect of ablation of a single unit on network performance reported negative relationships between selectivity index and loss (Morcos et al., 2018). However, the findings in that study were mainly based on the units in layers closer to the input layers, while the negative relationship was not observed in the higher layers. The contrasting findings in the present study may relate to differences in the network architectures and/or training methods. More importantly, although we found negative relationships, we highlight the importance of the selective unit, because better performance was obtained if the networks contained both selective and non-selective units.

Present results suggest a possible architecture for limited computational resources. DCNNs can be installed onto a variety of devices including smartphones. However, a problem with installing DCNNs on such terminals or edge devices is the limitation of computation resources. A possible way to overcome this limitation is to downsize DCNNs by decreasing the number of units (Anwar et al., 2015; Li et al., 2016; Molchanov et al., 2016). Our results suggest that removing units based on their degrees of stimulus selectivity may provide an efficient way for downsizing DCNNs.

**Figure Legends**

**Figure 1.** Stimulus selectivity of units in deep convolutional neural networks (DCNNs). *A,* Normalized response amplitudes of a stimulus selective (top) and non-selective units in fc6 of Alexnet. Stimuli were sorted according to their response magnitudes. *B,* Frequency distributions of the category selectivity index (CSI) of units in fc6 (left) and fc7 (right) of Alexnet. *C,* Frequency distributions of sparseness of units in fc6 (left) and fc7 (right) of Alexnet. *D,* Comparisons of stimulus selectivity across layers of Alexnet evaluated with CSI (top) and sparseness (bottom). In the box plot, the center of each box (black horizontal line) is the median, and the top and bottom of the box are the upper and lower quartiles, respectively. The attached whiskers connect the most extreme values within 150% of the interquartile range from the end of each box.

**Figure 2.** Frequency distributions of the loss_ratio after deletion of one unit in the DCNNs. *A,* Frequency distributions of the loss_ratio for fc6 (left) and fc7 (right) of Alexnet. *B,* Frequency distributions of the loss_ratio for fc6 (left) and fc7 (right) of VGG-19.

**Figure 3.** Relationships between CSI of single unit and the loss_ratio after deletion of the corresponding single unit in DCNNs. *A,* fc6 (left) and fc7 (right) of Alexnet. *B,* fc6 (left) and fc7 (right) of VGG-19. In each graph, 4,096 points derived from 4096 units are plotted.

**Figure 4.** Comparisons of correlation coefficients between stimulus selectivity and the loss_ratio across layers. *A*, Alexnet. *B*, VGG-19. Stimulus selectivity was quantified with CSI (red, solid lines with open circles) and sparseness (blue, broken lines with asterisks). Horizontal dotted line is at the correlation coefficient of zero.

**Figure 5.** Comparisons of the loss_ratio between the related category and other categories. Here, the related category is the category in which the optimal-stimulus image of the deleted unit belongs. *A*, fc6 of Alexnet. *B*, fc6 of VGG-19. Horizontal dotted line is at the loss_ratio of zero.

**Figure 6.** Relationships between correct rate and proportions of the non-selective unit. Stimulus selectivity was evaluated with CSI. The number of units was 1,000. Each line represents data from a single trial, with a total of 50 trials performed. *A*, fc6 (left) and fc7 (right) of Alexnet. *B*, fc6 (left) and fc7 (right) of VGG-19.

Table 1, Probability assigned to the correct label, top-5 accuracy and top-1accuracy of Alexnet and VGG-19 to the all images in ILSVRC 2012 validation set (all, 50,000 images) and selected images (selected, 5,000 images).

| Model | Stimulus set | Probability | Top-5 | Top-1 |
|---|---|---|---|---|
| Alexnet | all | 0.4599±0.3918 | 0.7951 | 0.5645 |
| | selected | 0.9361±0.1266 | 1.0000 | 0.9942 |
| VGG-19 | all | 0.5608±0.3914 | 0.8683 | 0.6590 |
| | selected | 0.8862±0.2050 | 0.9962 | 0.9554 |

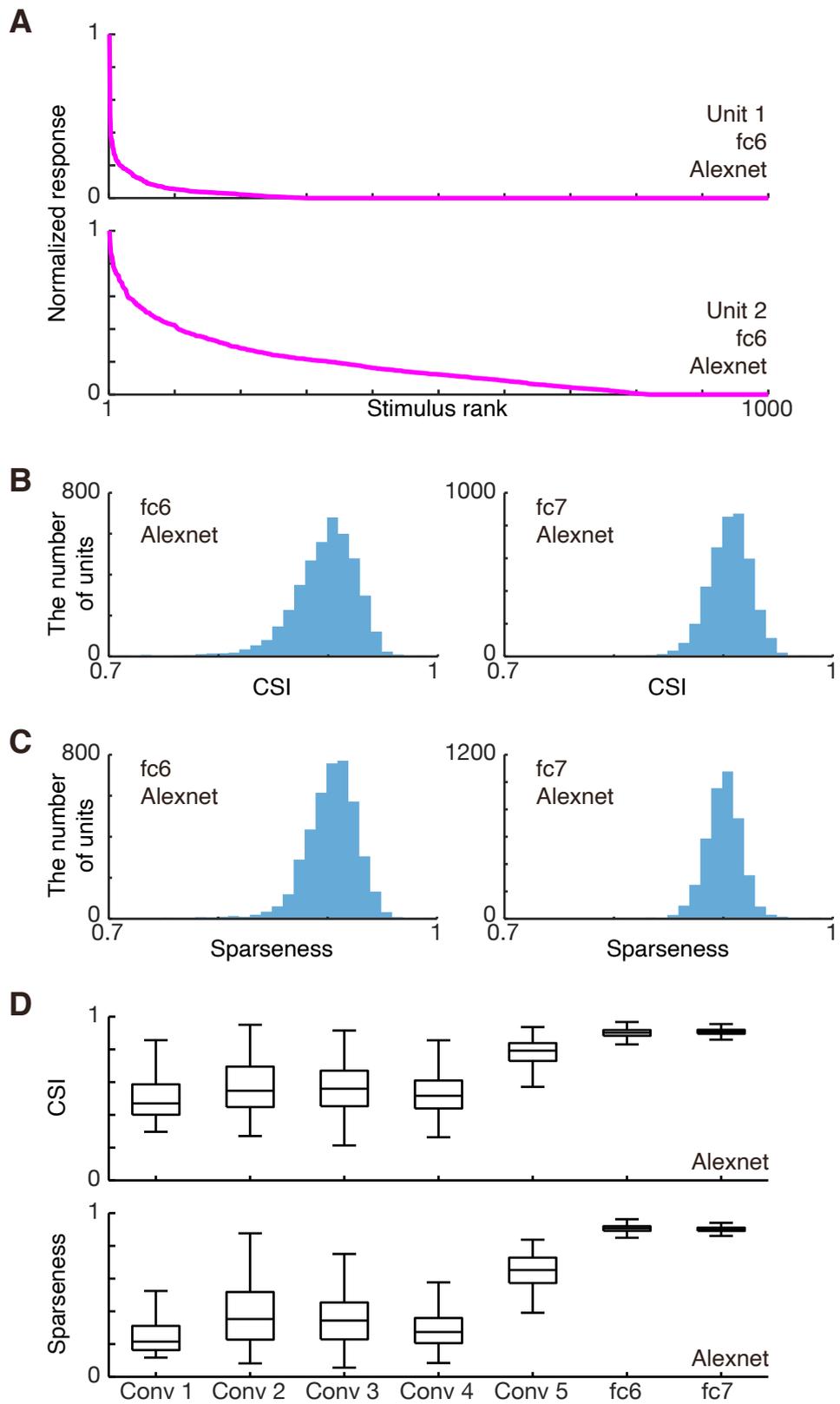

Figure 1
Kanda et al.

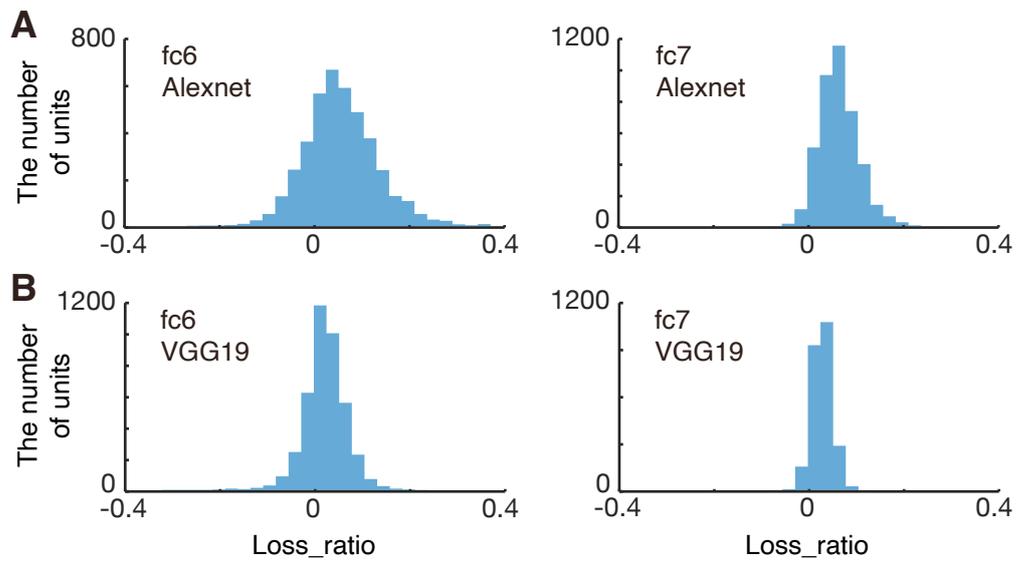

Figure 2
Kanda et al.

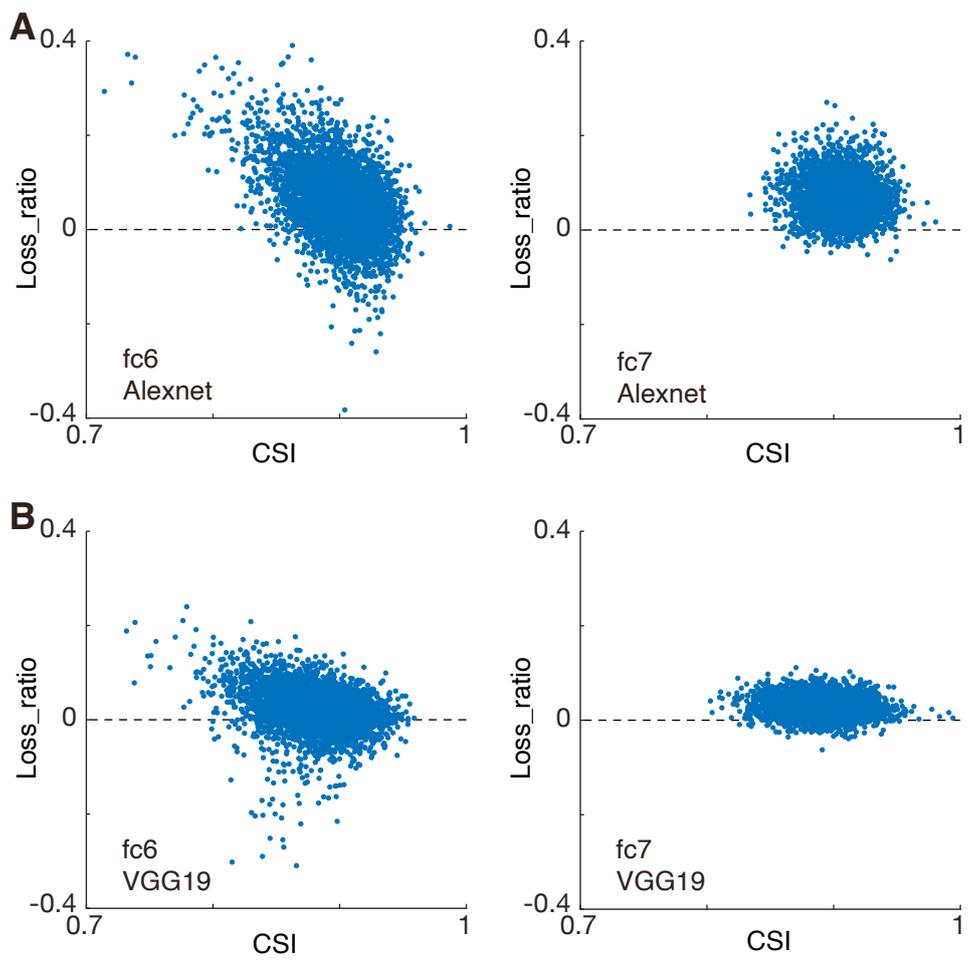

**A**

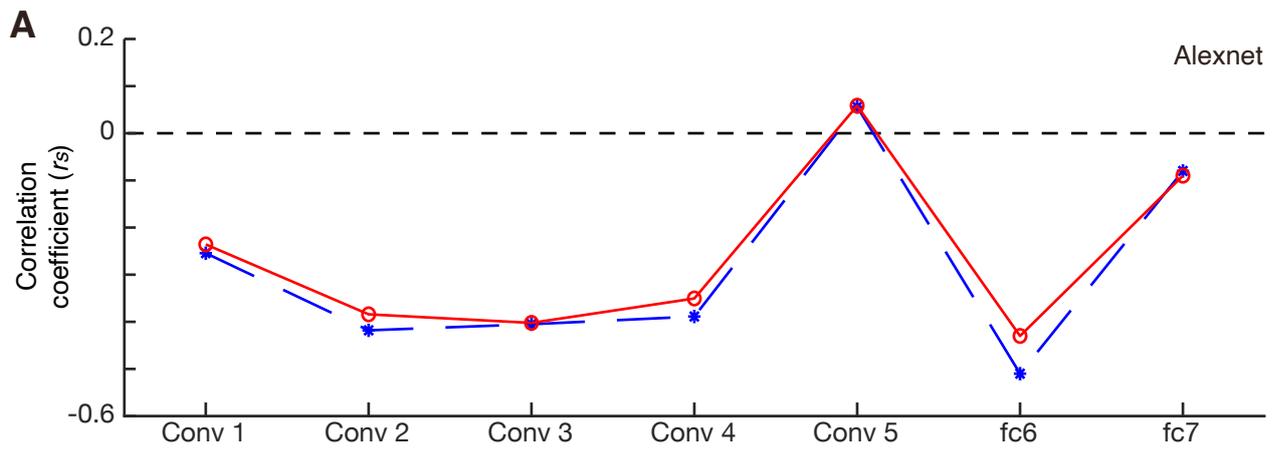

**B**

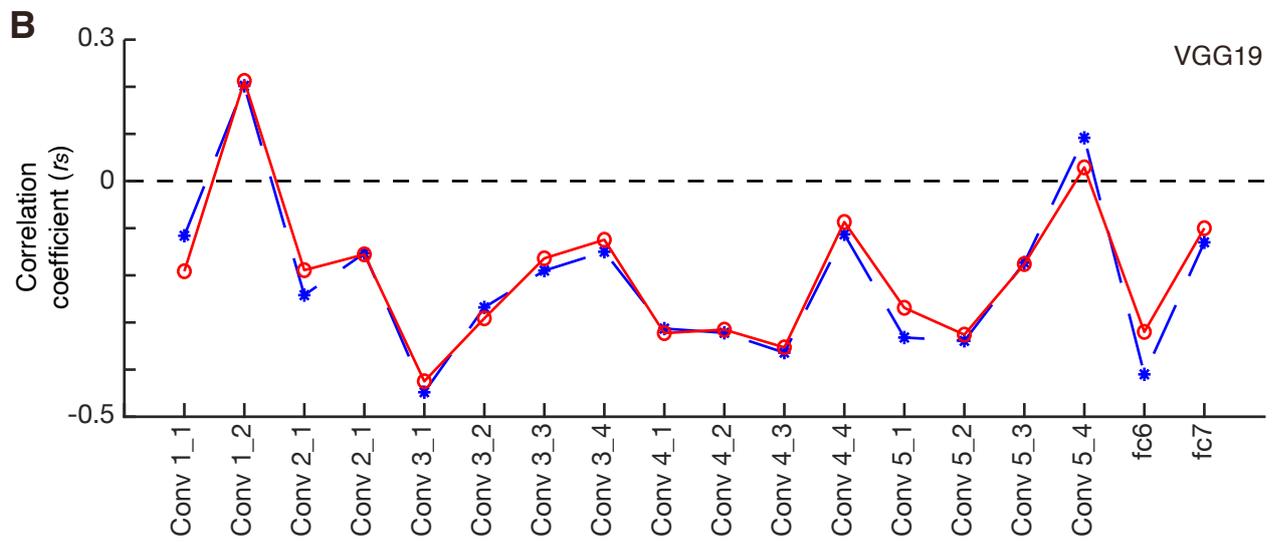

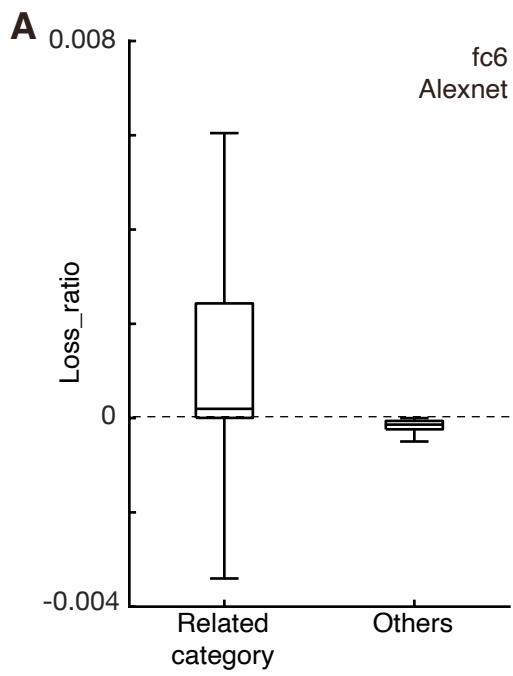

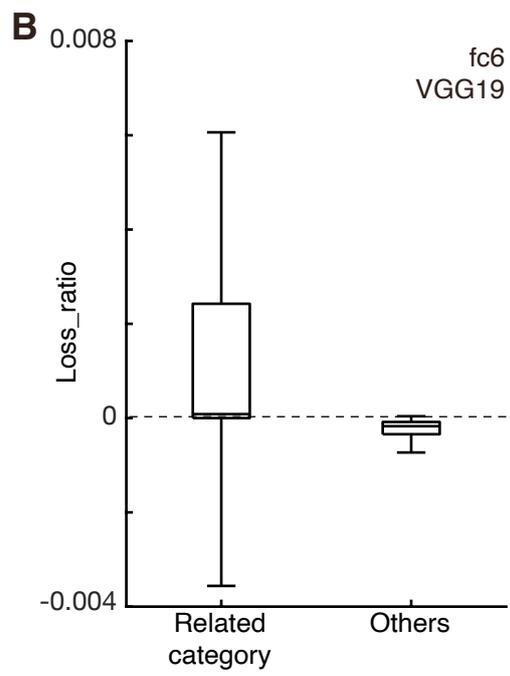

Figure 5
Kanda et al.

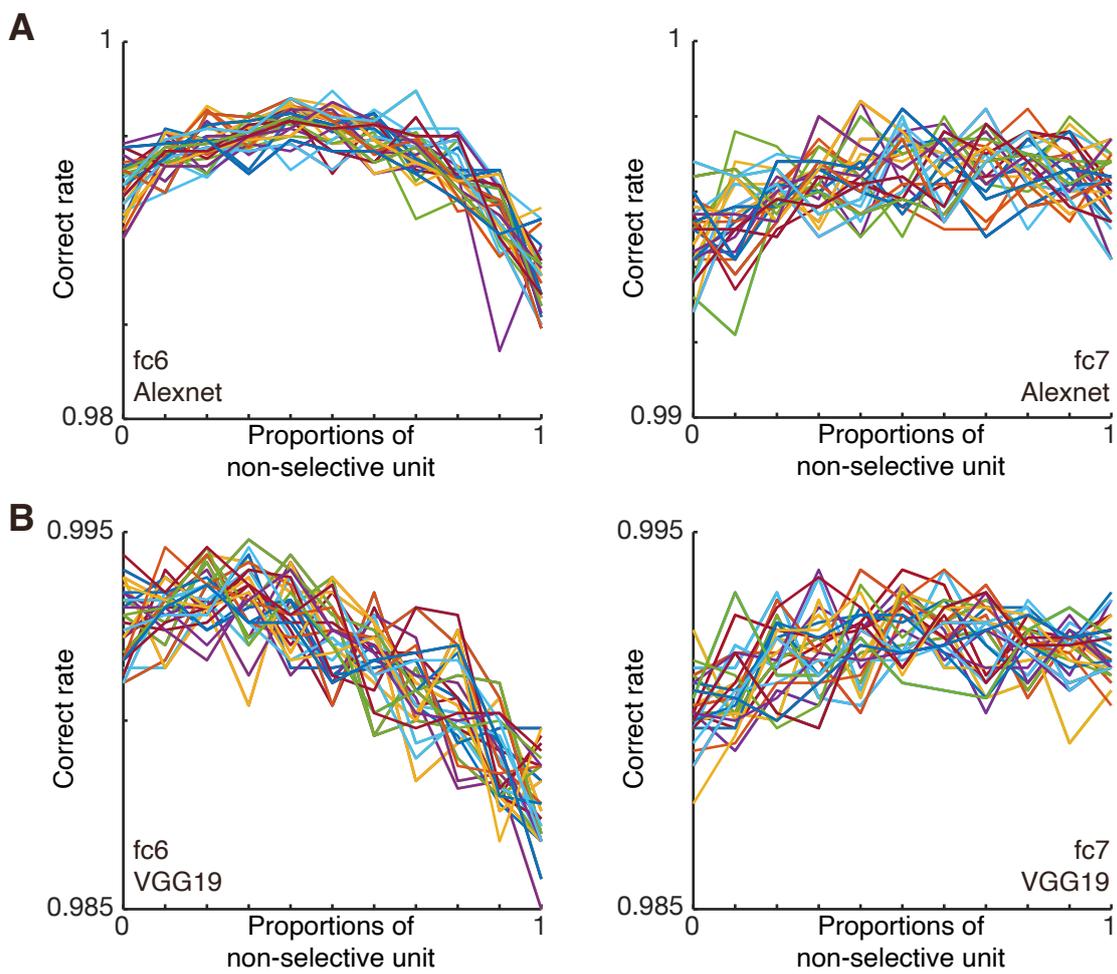